# VARIANTS OF INTRINSIC DISORDER IN THE HUMAN PROTEOME AND DISEASES

Antonio Deiana[1] and Andrea Giansanti*[1,2]

[1]Physics Department, Sapienza University of Rome, P.le A. Moro 2, 00185 Rome, Italy.
[2]INFN, section Roma1, P.le A. Moro 2, 00185 Rome, Italy.

* Correspondence to: andrea.giansanti@roma1.infn.it

**Abstract**

In this paper we propose a straightforward operational definition of variants of disordered proteins, taking the human proteome as a case study. The focus is on a distinction between mostly unstructured proteins and proteins which contain long unstructured regions accommodated in an overall folded structure. Two parameters control our distinction: $dr$, the percentage of disordered residues and $L_d$, the length of the shortest disordered domain in the sequence. In particular we distinguish: i) **Not disordered proteins (NDPs),** that either have all their residues ordered or do not have disordered segments longer than 30 residues ($L_d < 30$) nor more than 30% of disordered residues ($dr < 30\%$); ii) **Proteins with intrinsically disordered regions (IDRPs),** that have at least one disordered domain longer than 30 residues ($L_d \geq 30$), but disordered in less than 30% of their residues ($dr < 30\%$); iii) **Proteins that are intrinsically disordered (IDPs),** that have both at least one disordered segment longer than 30 residues ($L_d \geq 30$) and that are disordered on more than 30% of their residues ($dr \geq 30\%$); iv) **Proteins with fragmented disorder (FRAG_IDPs),** that do not have a disordered fragment longer than 30 residues ($L_d < 30$) but that, nevertheless, have at least 30% or more of their residues predicted as disordered ($dr \geq 30\%$). The potential use of these variants is checked over several groups of disease-related proteins. Our main conclusions point out that IDRPs are more similar to NDPs than to IDPs. IDRPs and NDPs have a similar functional repertoire and probably share a lock-and-key mechanism of interaction with substrates. IDRPs and IDPs are differently present among human disease-related proteins. IDRPs probably do not play a specific role in the development of complex diseases, since their frequency is similar in disease-related proteins and in the entire human proteome. IDPs can play a role in the emergence of cancer, neurodegenerative, thyroid and liver disease. The specific relevance we observe of IDRPs in diseases related to calcium homeostasis is an exception that deserves further investigation. As further investigation deserve the few human proteins with extended disorder, that we have called here FRAG_IDPs.

**Keywords:**

## 1. Introduction

Since the mid-1990s, evidences and the concept of structurally disordered but functional proteins massively emerged as a new theme in protein science [1-3]. The idea of an intrinsically disordered (natively unfolded, conformationally heterogeneous) protein is that of a polypeptide which samples broad distributions of conformational parameters (e.g., gyration radii [4, 5]). The opposite view is that of a globular (ordered) protein, as a solid-like polymer in which diffusive motions are quite limited and which is, thus, prone to be folded into a three-dimensional structure where each atom has a well defined average equilibrium position. Many have used to call this new trend in protein science as the shift of a paradigm [6]: despite the absence of a well-defined three-dimensional conformation, intrinsically disordered proteins are involved in important cellular functions, such as regulation and signaling [7-12], alternative splicing [13-15], as well as in the development of many diseases [16-18].

The focus of this paper is on the distinction, among intrinsically disordered proteins of two variants: IDPs and IDRPs, as specified below. The potential use of this distinction in protein science is illustrated by investigating how the two variants are distributed in several groups of disease-related proteins.

The distinction between a fully or a mostly unstructured protein (IDP) and a protein which contains unstructured regions (IDRP) is quite well established (see e.g. [2, 19]).

On the one hand, following the current literature, an intrinsically disordered protein (IDP) is a protein that lacks a three-dimensional structure in a long portion of its polypeptide chain. IDPs do not interact with the substrates through a lock-and-key mechanism and can interact with many substrates with high specificity and low affinity [6, 20].

On the other hand, the existence of mostly folded proteins but with intrinsically disordered regions (IDRPs) has been extensively discussed in the literature [21, 22]. These proteins have a tertiary structure and therefore it is possible to conceive that they still interact with substrates through a lock-and-key mechanism. The presence of loops or domains without three-dimensional conformation (e.g. linear motifs and Molecular Recognition Features (MoRFs)) can be important for IDRPs to target low affinity substrates, enlarging the repertoire of protein interactions [11].

In the majority of the papers in the field, IDPs have been identified as proteins with at least one long disordered segment, i.e. a segment of at least 30 consecutive disordered amino acids. It has been reported that about 33% of proteins from Eukarya have a long disordered segment [23]. It has been also reported that 79% of cancer-related proteins and 57% of proteins related to cardio-vascular diseases [24, 25] are intrinsically disordered, suggesting but not proving a key role of protein disorder in the development of many human diseases [16, 17]. Of course, the presence of long disordered segments does not imply that a protein is intrinsically disordered, i.e. that it lacks a stable overall three-dimensional conformation. The distinction between IDPs and IDRPs depends on the relative proportion of the unstructured regions with respect to the rest of the protein.

In this work we operationally revisited this distinction, taking into account the overall percentage of disorder *dr* detected in the sequence. It has been shown that proteins with a percentage *dr* of disordered residues larger than 30% have a higher turn-over with respect to other proteins [26]. Therefore, the overall percentage of disorder *dr* seems to be an important parameter to distinguish variants of intrinsic disorder. This suggests the criterion that we propose in this paper: a protein is intrinsically disordered (IDP) if it has at least one disordered segment of length $L_d > 30$ and more than 30% of its amino acids predicted as disordered. This criterion is indeed not completely new (see, e.g. [15, 27]). By contrast, we label here as IDRPs proteins with at least one disordered segment of length $L_d > 30$ but with a percentage of disorder *dr* lower than 30%. We mentioned also another very limited group of human proteins, characterized by a kind of fragmented disorder, which is not organized into domains.

We found that 26% and 36% of human proteins are IDRPs and IDPs, respectively. IDPs are rich in hydrophilic and charged amino acids and poor in hydrophobic, apolar, bulky amino acids. In a hydropathy-charge plot, they have a lower hydrophobicity than folded non-disordered proteins (NDPs), consistently with previous results in the literature [18, 28]. Interestingly, IDRPs have values of hydrophobicity and net charge more similar to those of non disordered, globular proteins than to those of IDPs (see figure 1 below). This key observation points to the fact that, on physical-chemical properties, IDRPs are more similar to folded proteins than to unfolded ones. Moreover, IDRPs have ontologies and functional profiles that are closer to those of non disordered proteins than to those of IDPs, as shown in section 3.3. We concluded that IDRPs and IDPs are distinct variants of disordered proteins and therefore they must be considered separately in their functional roles and as agents in the emergence of diseases.

We considered different pathologies related to: calcium homeostasis, cancer, cardio-vascular diseases, diabetes, genetic, liver, neurodegenerative and thyroid diseases. We observed that the incidence of IDRPs among the proteins related to the above mentioned diseases is about 30%, slightly higher than in the whole human proteome (i.e. 26%). This suggests that the category of overall folded proteins with long disordered domains, which we call here IDRPs, must be important to enrich the repertoire of interactions of human proteins, but probably does not specifically contribute to the development of diseases.

At variance with IDRPs, the incidence of IDPs in each group of diseases is quite diversified, with a prevalence of IDPs among proteins related to cancer, neurodegenerative, liver and thyroid diseases and a prevalence of non-disordered proteins (NDPs) in cardio-vascular diseases and diabetes.

Our observations suggest that statements arguing generic causative involvements of IDPs in diseases, though understandable in the initial phases of the study of IDPs, nowadays must be considered with caution. IDRPs and IDPs are distinct variants of disorder and, in our opinion, they must be investigated separately. Moreover, IDRPs do not contribute specifically to the development of diseases. In regard to IDPs, they seem to be involved in the development of cancer, neuro-degenerative diseases and diseases of the liver and thyroid, but not in cardio-vascular diseases and diabetes that instead seem to be controlled by ordered proteins. In our opinion, statements generically suggesting that protein disorder is involved in the development of diseases are probably wrong, since we have shown that there are diseases associated to non-disordered proteins. However, their clear involvement in several diseases indicates that IDPs as specific agents deserves further investigations, focussed on specific pathologies.

## 2. Methods

**2.1 Protein sequences.**
We downloaded the human proteome from the Uniprot – Swissprot database (http://www.uniprot.org/uniprot), release of September 2016 [29]. We selected human proteins by searching for reviewed proteins belonging to the organism Homo Sapiens id. 9606. Of the 20177 protein sequences, 60 were discarded since they contain atypical amino acid symbols (B, U, Z, X). We considered, then, 20117 protein sequences.
We identified disease-related proteins by searching for human proteins annotated with the keywords available at: ftp://aglab.phys.uniroma1.it/pub/ .

**2.2 Disorder prediction**
The methods of the PONDR VSL2 family are support vector machines trained to recognize disordered residues from the amino acid composition of the protein sequence they are embedded in [30, 31]. We used the VSL2B version, which was trained on 1327 ordered and disordered proteins, selected from the Protein Data Bank [32] and from DisProt [33]. The training set comprised proteins with experimentally identified disordered domains longer than 30 residues and proteins with short disordered domains. To control overestimation of disorder prediction we retuned VSL2B parameters to match *sensitivity* (TP/(TP+FN)) with *precision* (TP/(TP+FP)), following the protocol in [34].

**2.3 Classification of disordered proteins in the human proteome: combining two criteria.**

We partitioned the human proteome into four variants of protein sequences, namely:

1. **Not disordered proteins (NDPs),** that either have all their residues ordered or do not have disordered segments longer than 30 residues ($L_d < 30$) nor more than 30% of disordered residues ($dr < 30\%$);
2. **Proteins with intrinsically disordered regions (IDRPs),** that have at least one disordered domain longer than 30 residues ($L_d \geq 30$), but disordered in less than 30% of their residues ($dr < 30\%$);
3. **Proteins that are intrinsically disordered (IDPs),** that have both at least one disordered segment longer than 30 residues ($L_d \geq 30$) and that are disordered on more than 30% of their residues ($dr \geq 30\%$);
4. **Proteins with fragmented disorder (FRAG_IDPs),** that do not have a disordered fragment longer than 30 residues ($L_d < 30$) but that, nevertheless, have at least 30% or more of their residues predicted as disordered ($dr \geq 30\%$).

NDPs are, clearly, proteins with a limited number of disordered residues and absence of disordered domains. IDRPs are protein sequences with regions that are locally disordered, characterized by the presence of disordered domains in a structure that might well be globally folded. IDPs are the true intrinsically disordered proteins, with sequences that not only embed disordered domains, but which are also disordered in a relevant percentage of their residues. FRAG_IDPs, are a small number of sequences characterized by highly distributed disorder.

## 2.4 Charge-hydropathy (CH) plots.

One of the earliest binary classifier of disordered proteins has been the charge-hydropathy (CH) plot, based on the Kyte-Doolittle normalized scale. The discriminating line we used here is the original one: <charge> = 2.785 x <hydrophobicity> - 1.151 [28].

## 2.6 Functional screening and ontologies

We attributed the four variants of proteins NDPs, IDRPs, IDPs and FRAG_IDPs to several protein classes and ontologies, following the annotations of the Gene Ontology Consortium: molecular functions, biological processes and cellular components [35]. We used the PANTHER (Protein Annotation Through Evolutionary Relationship) classification scheme (http://www.pantherdb.org) [36, 37].

To evaluate the functional profiles of each variant of disordered proteins we estimated the conditional probability P(protein-class|variant): for each variant, we computed the number of proteins of that variant in each PANTHER class. This number is then divided by the total number of proteins in the specified variant.

To evaluate how a PANTHER protein class is enriched in a variant of disorder we estimated the conditional probability P(variant|protein-class): that is the fraction of proteins belonging to a variant of disorder among the proteins of a given class. More clearly, given a specific class, we computed the number of proteins in the class which belong to each one of the protein variants: NDPs, IDRPs, IDPs and FRAG_IDPs. This number is then divided by the number of proteins in the specific class.

## 3. Results

### 3.1. Variants of disorder in the human proteome

The 20117 sequences of human proteins were partitioned into the four variants: NDPs, IDRPs, IDPs and FRAG_IDPs (table 1). Note that IDRPs and IDPs amount to 62%. The percentage of IDPs

(36%) is consistent with previous estimates in the human proteome [15]. FRAG_IDPs, characterized by a sparse distribution of disorder, are represented by just a few cases worth to be further investigated, in future studies, both as a category and as single cases.

| Variant | N. of proteins | % |
|---|---|---|
| NDPs | 7423 | 37 |
| IDRPs | 5327 | 26 |
| IDPs | 7214 | 36 |
| FRAG_IDPs | 153 | 1 |

**Table 1. Percentages of ND, IDRs, IDPs, FRAG_IDPs in the human proteome.**
Disordered proteins, IDRPs and IDPs, constitute 62% of the human proteome, consistently with previous estimates. The remaining 38% of the human proteins have been classified either as not disordered (NDPs) or as FRAG_IDPs.

### 3.2. Charge-hydropathy (CH) plots.

Charge-hydropathy (CH) plots are basic tools to separate disordered from ordered protein sequences [28]. Disordered protein sequences are indeed characterized by low hydrophobicity and high net charge [28, 38]. Clearly the mass of NDPs and IDRPs, that in many previous works have been classified as disordered, mainly belongs to the ordered (low net-charge and low hydropathy) phase, whereas IDPs and FRAG_IDPs are spread on both sides of the separating line. This observation points out that IDRPs (with at least a disordered domain longer than 30 residues but disordered in less than 30% of their residues) have physical-chemical properties more similar to folded proteins than to unfolded ones, so caution is required in labeling these proteins as disordered, as it has been done for long time in the literature, since it can be misleading. The interaction of IDRPs with substrates probably is of the lock-and-key type, quite different by the high-affinity, low-specificity interactions typical of IDPs.

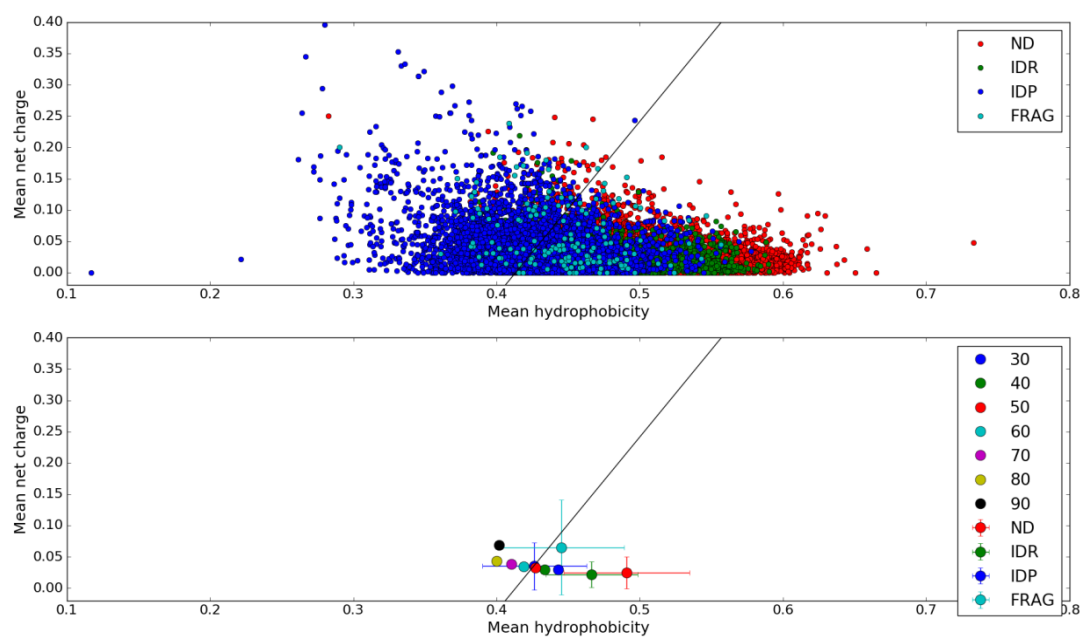

**Figure 1. Variants of disorder in human proteins represented on the CH plot**.

The solid line is the separating line (<net charge> = 2.785 x <hydrophaty> - 1.151.) based on the Kyte-Doolittle scale [28]. A) IDRPs are closer to NDPs than to IDPs and FRAG_IDPs. B) The centroids of NDPs, IDRPs IDPs and FRAG_IDPs are shown; IDRPs, that in many studies have been classified among disordered proteins, lay evidently in the ordered region of the plane. Moreover, percentiles of IDPs - characterized by the % amount of disordered residues - are also represented. It is worth noting that the centroids of IDPs and FRAG_IDPs lay in the ordered region but with great overlaps with the other side of the line (figure 1B). The driving parameter that pushes towards the disordered region is clearly the overall percentage of disorder in the sequence. In particular, only IDPs with at least 60, 70, 80 and 90% of predicted disordered residues lay in the disordered region of the plane. Error bars are root mean square deviations.

Noteworthy, the percentage of disorder is an important parameter that drives the representative point of IDPs across the separating line of the CH plane. In particular, if the percentage of disorder is less than 50% then the centroid is in the ordered phase, whereas it crosses the separating line if the percentage of disordered residues is more than 60%. This would suggest a further re-definition of IDPs, using a critical value of *dr* not less than 50%.
From the graph in figure 1B it is possible to straightforwardly evaluate overlaps between the variants of disorder introduced in this paper. To each centroid in the CH plot of figure 1B is associated a rectangle constructed over the standard deviation bars; then the overlap between two centroids is evaluated as the intersection area of the corresponding rectangles. The overlaps between the centroids of disorder variants in the CH plane are then collected into an overlap matrix (table 2). Note that also in this representation, based on physical-chemical parameters such as charge and hydrophobicity, IDRPs have a bigger overlap with NDPs (2.2) than with IDPs (1.3). This indicates that IDRPs are a variant of proteins with localized disordered domains that do not substantially alter the overall globular architecture, typical of folded proteins. In the CH plot FRAG_IDPs, have a considerable overlap both with NDPs and IDRPs, but their overlap with IDPs is almost double, suggesting that these proteins with fragmented disorder are a sub-variant of IDPs.

|  | NDPs | IDRPs | IDPs | FRAG_IDPs |
|---|---|---|---|---|
| NDPs | - | 2.2 | 0.8 | 2.6 |
| IDRPs |  | - | 1.3 | 3.1 |
| IDPs |  |  | - | 5.3 |
| FRAG_IDPs |  |  |  | - |

**Table 2. Overlaps between ND, IDRS, IDPs and FRAG_IDPs on the CH plot.**
Based on the centroids of figure 1B it is possible to compute relative overlaps of variants of disorder in the CH plane. Note that, consistently with table 1, IDRPs have a bigger overlap (in arbitrary units) with NDPs than with IDPs. FRAG_IDPs have a quite peculiar location in the CH plane that give them a considerable overlap with both NDPs and IDRPs, nevertheless their biggest overlap is with IDPs (5.3).

### 3.3 Distribution of ND, IDRs, IDPs and FRAG_IDPs over PANTHER protein classes and gene ontologies.

We projected human proteins belonging to the four variants of disorder over the PANTHER classification scheme [36, 37]: protein-classes and gene-ontology terms. In figure 2 histograms of the conditional probability P(protein-class|variant) are given, for the four variants of disorder: NDP, IDRP, IDP and FRAG_IDP. This is the probability (estimated as relative frequency) of a PANTHER protein-class, among the proteins belonging to any of the four variants of disorder. It is worth to note that graphs in figure 2 do not represent the enrichment of the different PANTHER protein classes in the variants of disorder; they represent – for each variant – its projection over the classes, the fractions of human protein sequences that, in each variant of disorder, belong to the different PANTHER protein-classes. Consider as a relevant case that of IDPs (blue columns): this variant is particularly projected on the *nucleic acid binding* (PC00171) class and on *transcription factor* (PC00218). The few FRAG_IDPs (cyan columns) are highly projected on the *nucleic acid binding* (PC00171) class as well, on the *transporter* class (PC00227) and, on the same footing with other variants, on the *enzyme modulator* (PC00095) class. By contrast, NDPs and IDRPs are more uniformly spread over the different protein-classes. This observation clearly suggests that NDPs and IDRPs on one side and IDPs and the few FRAG_IDPs on the other have different functional repertoires; more generic those of NDPs and IDRPs, but more focused those of IDPs and FRAG_IDPs.

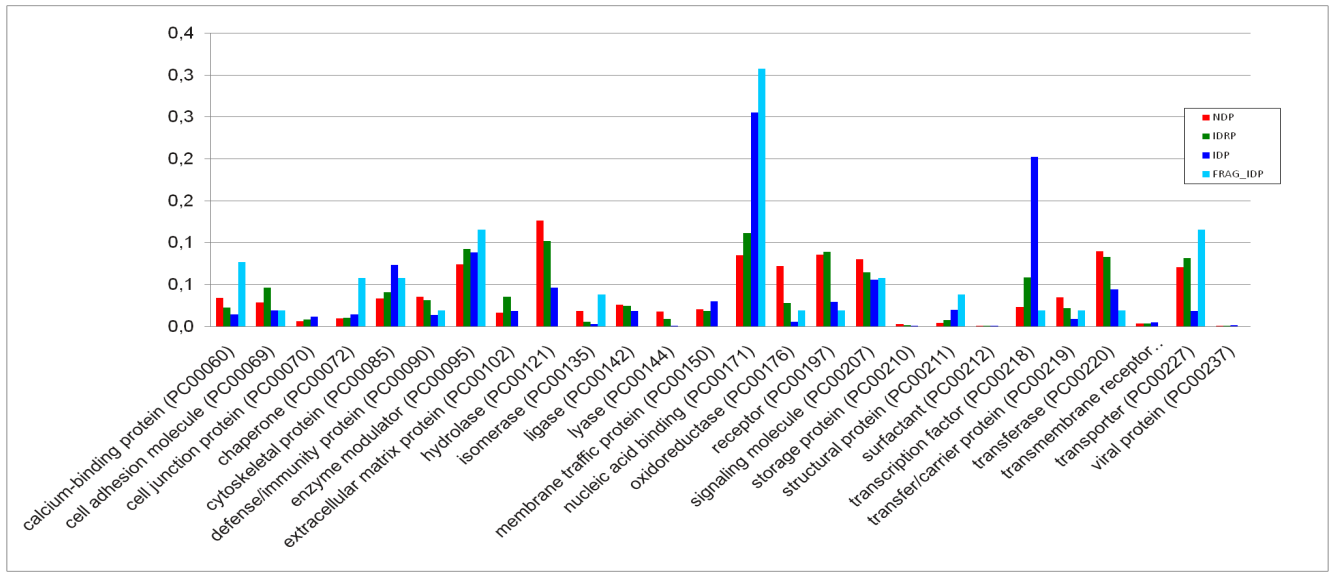

**Figure 2. NDPs, IDRPs, IDPs and FRAG_IDPs projected over the space of PANTHER protein classes.** This histogram then represents a graph of P(protein-class|variant), measuring how much each variant of disorder is projected over the different classes.

To give a quantitative measure of similarities between functional repertoires of the different variants we have associated probability vectors such as: P(protein-class|variant), P(biological-function|variant), P(molecular-process|variant) and P(cellular-component|variant) to each variant of disorder, then a distance between the variants has been computed using the Euclidean distance between the corresponding unit vectors. The resulting distance matrices are collected in table 3.

| | | NDP | IDRP | IDP | FRAG | | | NDP | IDRP | IDP | FRAG |
|---|---|---|---|---|---|---|---|---|---|---|---|
| Protein-Class | NDP | 0.000 | 0.08 | 0.29 | 0.30 | Molecular function | NDP | 0.00 | 0.12 | 0.35 | 0.31 |
| | IDRP | | 0.00 | 0.24 | 0.27 | | IDRP | | 0.00 | 0.25 | 0.26 |
| | IDP | | | 0.00 | 0.24 | | IDP | | | 0.00 | 0.28 |
| | FRAG | | | | 0.00 | | FRAG | | | | 0.00 |

| | | NDP | IDRP | IDP | FRAG | | | NDP | IDRP | IDP | FRAG |
|---|---|---|---|---|---|---|---|---|---|---|---|
| Biological Process | NDP | 0.00 | 0.07 | 0.11 | 0.13 | Cellular component | ND | 0.00 | 0.09 | 0.23 | 0.17 |
| | IDRP | | 0.00 | 0.07 | 0.10 | | IDR | | 0.00 | 0.15 | 0.11 |
| | IDP | | | 0.00 | 0.13 | | IDP | | | 0.00 | 0.12 |
| | FRAG | | | | 0.000 | | FRAG | | | | 0.000 |

**Table 3 Relative distances in the space of PANTHER protein-classes and gene ontology categories.** The four variants of disorder in the human proteome are projected over PANTHER protein-classes and gene ontology terms under the three categories: biological process, molecular function and cellular component. Distances are computed as euclidean distances between representative probability vectors associated to each variant of disorder.

Globally, these distance matrices clearly show that the functional repertoires of IDRPs are closer to those of NDPs than to those of IDPs and FRAG_IDPs, indicating that the distinction we are introducing here is not trivial: IDPs are indeed distinct from IDRs, as we shall illustrate, using other observables, in the following.

**3.4 Enrichment in variants of disorder of the different PANTHER protein-classes.**

To evaluate how each PANTHER protein-class is enriched in one or more of the four variants, we produced histograms representing conditional probabilities such as P(disorder-variant|protein-class), that is to compute, among the proteins that belong to a specific class, the fraction of each variant (see figure 3).

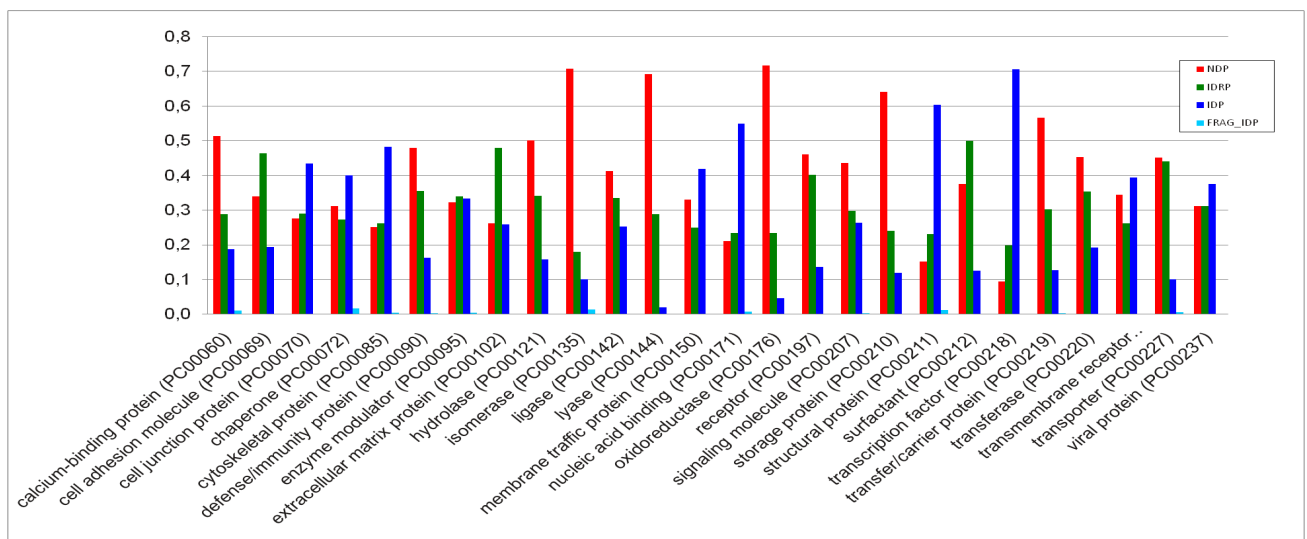

**Figure 3. NDPs, IDRPs, IDPs and FRAG_IDPs projected over the space of PANTHER protein classes.** This histogram, complementary to that in the previous figure 2, represents the conditional probabilities P(variant|protein-class); the proteins in each one of the 29 PANTHER protein-classes are partitioned into the four variants of disorder; the normalization is over each protein-class. In this histogram, clearly FRAG_IDPs are not visible, due to their limited occurrence (see table 1)

Interestingly, IDPs clearly dominate among *cytoskeletal proteins* (PC0085), among *nucleic acid binding* (PC00171), *structural proteins* (PC00211), *transcription factors* (PC00218); they are possibly slightly overrepresented among *cell junction proteins* (PC00070), *chaperones* (PC00072), *membrane traffic proteins* (PC000150) and *viral proteins* (PC00237). NDPs dominate among *calcium binding proteins* (PC00060), *defense/immunity proteins* (PC00090), *transfer/carrier proteins* (PC00219) and, as expected, among enzymes: *hydrolases* (PC00121), *isomerases* (PC00135), *lyases* (PC00144), *oxidoreductases* (PC00176), *phosphatases* (PC00181), *proteases* (PC00190) *transferases* (PC00220). From data in figure 3 it is possible to separate PANTHER protein-classes on the basis of the variant of disorder in which they are enriched. The above sparse observations are collected in table 4, where, into separate columns, are collected the PANTHER protein classes that are enriched in different variants. As a first hand observation, note that protein classes dominated by IDPs refer to structural proteins such as PC00211 (extracellular structural proteins, e.g. involved in skin appendages), or cytoskeletal proteins (PC0085), that, interestingly, appear as hardly structured proteins, as is well known.

| NDPs | IDRPs | IDP |
|---|---|---|
| *Calcium binding* (PC00060) | *Cell adhesion molecules* (PC00069) | *Cytoskeletal proteins* (PC0085) |
| *Defense/immunity* (PC00090) | | |
| *Hydrolases* (PC00121) | | *Nucleic acid binding* (PC00171) |
| *Isomerases* (PC00135) | *Extracellular matrix proteins* (PC00102) | |
| *Lyases* (PC00144) | | *Structural proteins* (PC00211) |
| *Oxidoreductases* (PC00176) | | |
| *Phosphatases* (PC00181) | *Kinases* (PC00121) | *Transcription factors* (PC00218) |
| *Proteases* (PC00190) | | |
| *Storage proteins* (PC00210) | *Transporter proteins* (PC00227) | |
| *Transferases* (PC00220) | | |
| *Transfer/carrier proteins* (PC00219) | | |
| *Transporter proteins* (PC00227) | | |

**Table 4. PANTHER protein classes that are enriched in variants of disorder.**
This table collects PANTHER protein classes which are enriched in one of the three variants: ND, IDRs and IDPs. FRAG_IDPs are not statistically relevant, due to their limited number.

**3.5 Disordered proteins and diseases**

We reconsider then, a widely discussed theme which can be synthetically expressed by the question: does disorder play a key role in the development of diseases? In many of the previous works [16-18], intrinsically disordered proteins are in most cases defined as proteins with $L_d > 30$. As pointed out above, however, this definition does not distinguish among IDPs and IDRPs, and we argued that IDPs and IDRPs must be considered as two protein variants with different physical-chemical and functional properties, mainly due to the different percentage of disorder they contain.

In figure 4 we show the fractions of IDRPs among several disease-related proteins, compared with the overall percentage of IDRPs in the proteome of Homo Sapiens, taken as a reference. It is relevant to note that the frequency of IDRPs in disease-related proteins is 30%, only slightly higher than in the whole human proteome, i.e. about 26%. The only exception is represented by proteins associated to calcium homeostasis diseases, among which IDRPs exceed 46%. The observed occurrences of IDRPs point to a non-specific role in the development of disease, with the remarkable case of the enrichment in IDRPs shown by proteins related to calcium homeostasis diseases, that deserve further investigation.

In figure 5 we report the frequency of IDPs in the human proteome and in disease-related proteins. As we can observe, with respect to the frequency of IDPs in the human proteome, the frequency of IDPs is higher among proteins related to cancer, neurodegenerative diseases and the diseases of liver and thyroid, but is lower in proteins associated to cardio-vascular diseases and diabetes. The incidence of IDPs among proteins associated to genetic diseases and in the human proteome is about the same. The prevalence of IDPs in several diseases suggests that the absence of a well-defined three-dimensional structure, typical of IDPs, can play a role in the development of several diseases, i.e. cancer, neurodegenerative, liver and thyroid diseases, as suggested also by previous publications. On the other hand, cardio-vascular diseases and diabetes are probably not specifically influenced by disorder, at variance with previous suggestion in the literature [16, 17].

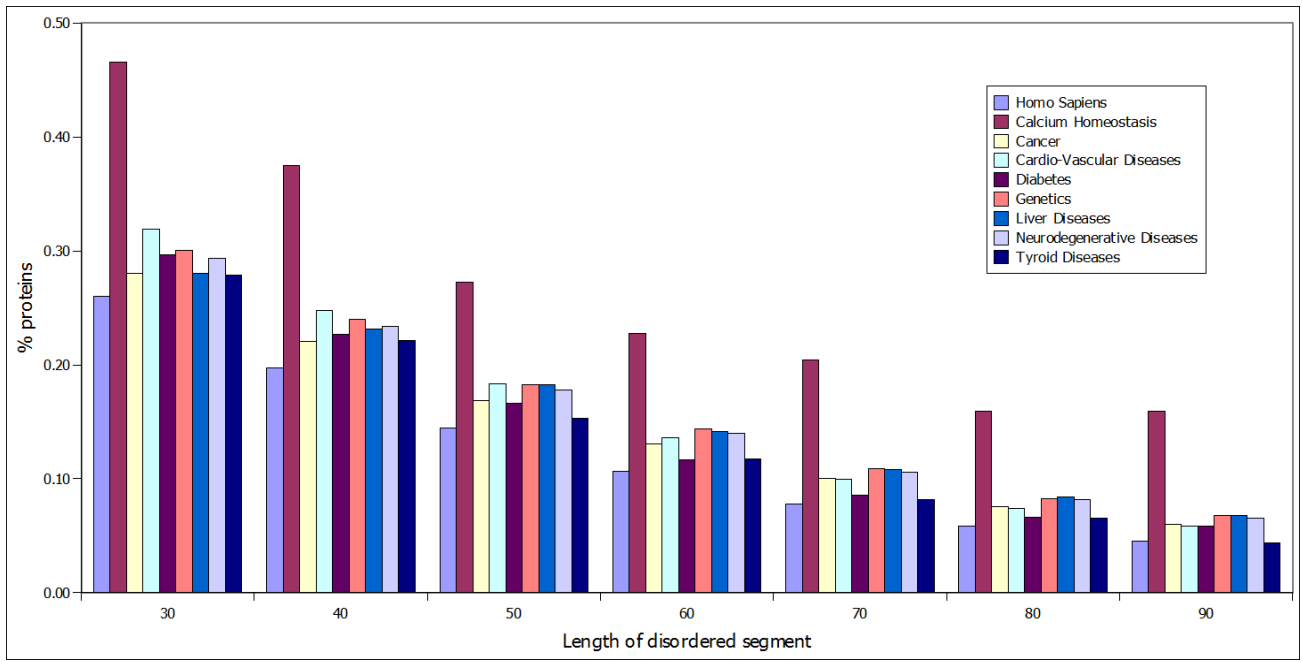

**Figure 4. Differential distribution of IDRPs in Homo Sapiens and in proteins annotated as disease-related, as a function of the minimal length of the disordered regions.**
IDRPs are proteins with $L_d \geq 30$ and $dr < 30\%$. The first group of bins contains IDRPs with at least one disordered segment with $L_d \geq 30$, the second group of bins contains IDRPs with at least one disordered segment with $L_d \geq 40$ and so on. The number of proteins in each bin, associated to a specific disease, is divided by the total number of human proteins associated to that disease. Clearly, the second group of bins contains all the proteins in the first group subtracted by those IDRPs that have disordered segments shortest than 40 residues, and so on. The requirement of longer and longer disordered stretches reduces the number of proteins fullfilling it, a trend observed in many papers (see e.g. [17], figure 4).

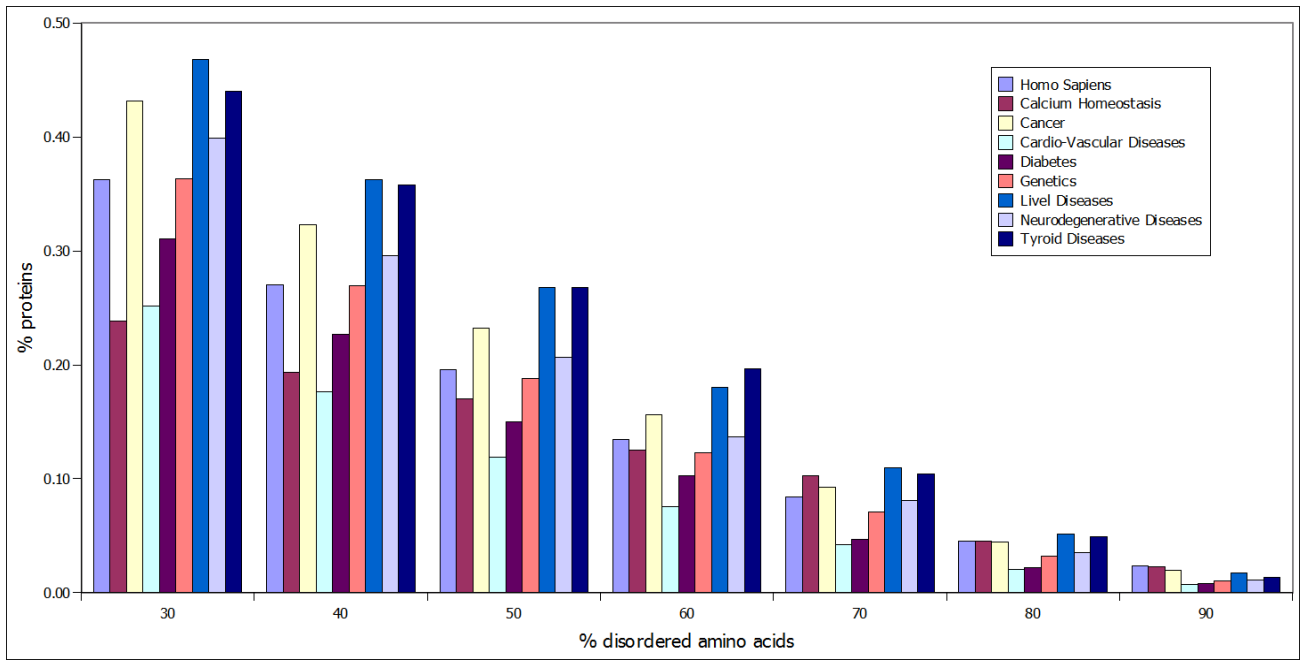

**Figure 5. Differential distribution of IDPs in the whole Homo Sapiens proteome and among disease-related proteins, as a function of the percentage of disorder.**
The first group of bins contains IDPs with at least 30% of disordered amino acids, the second group of bins contains IDPs with at least 40% of disordered amino acids and so on. The number of proteins in each bin, associated to a specific disease, is divided by the number of proteins in the human proteome associated to that disease. Clearly, the second group of bins contains all the proteins in the first group subtracted by those IDPs that have a percentage of disordered amino acids between 30% and 40% and so on.

It is important to point out that, from figure 5, we simply observe that IDPs are more frequent in several diseases with respect to the human proteome. This higher occurrence of IDPs in proteins associated to a disease however is not sufficient to conclude that disorder is important in the development of that disease, since we cannot exclude that the majority of proteins associated to that disease are not IDPs, although IDPs are more frequent than in the rest of the proteome. To investigate whether IDPs are the prevalent variant of proteins in a specific disease, we computed the conditional probability P(variant|disease-related) for each disease considered here, i.e. the number of NDPs, IDRPs, IDPs and FRAG_IDPs among the proteins associated to each disease, divided by the number of proteins associated to that disease. The result is reported in figure 6. IDPs are prevalent in cancer, liver, neurodegenerative and thyroid diseases. NDPs are prevalent among proteins related to cardio-vascular diseases and diabetes. IDRPs are prevalent in calcium homeostasis diseases. The frequency of NDPs, IDRPs and IDPs are similar in genetic diseases, with a slight prevalence of IDPs. This result suggests that the absence of a well-defined three-dimensional structure can play a role in the development of cancer and of neurodegenerative diseases, liver and thyroid diseases, whereas it does not play a specific role in the development of the other diseases. Interestingly, the relevance of IDRPs in diseases related to calcium homeostasis, suggests that these proteins, although well structured, should also have a certain flexibility, guaranteed by the presence of a long disordered loops, and this flexibility plays a role in the development of this kind of diseases

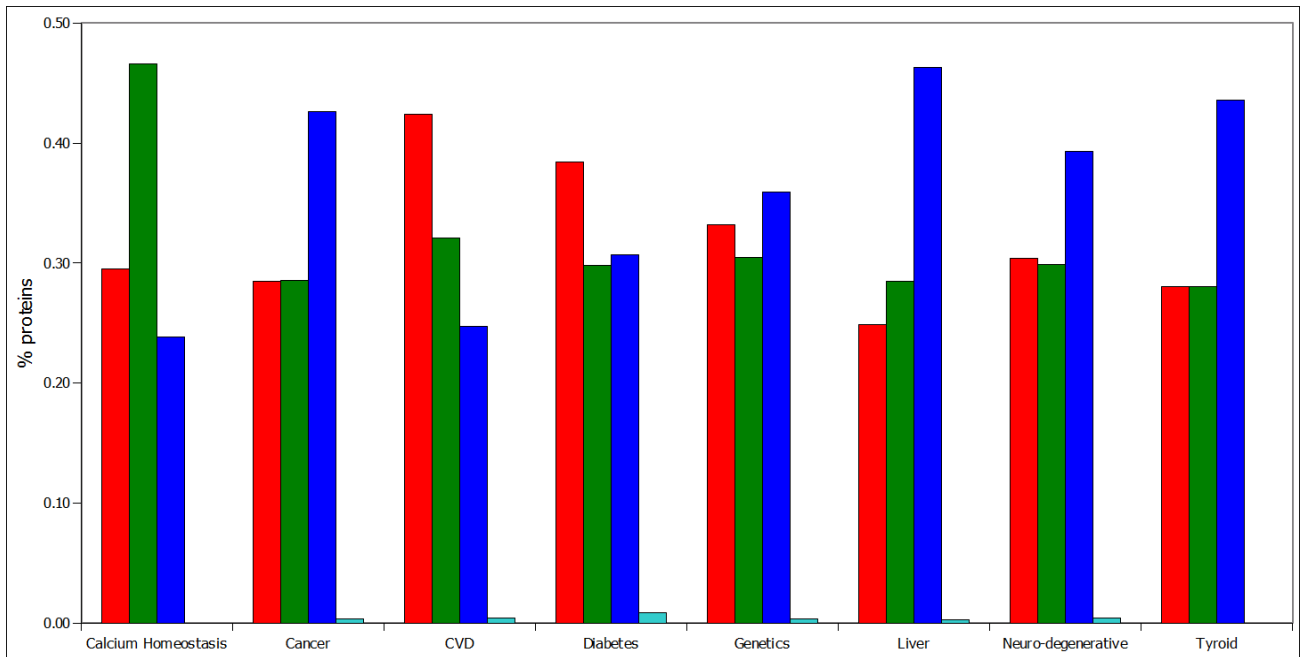

**Figure 6. Relative incidence of NDPs, IDRPs, IDPs and FRAG_IDPs among human proteins annotated as related to different diseases.**
For each group of disease-related proteins the sum of the protein frequencies is equal to 1. IDPs are relevant in cancer, liver, neurodegenerative and thyroid diseases. NDPs are prevalent in cardio-vascular diseases and diabetes. IDRPs are prevalent in calcium homeostasis diseases. The frequency of NDPs, IDRPs and IDPs are similar in genetic diseases, with a slight prevalence of IDPs.

**Discussion**

The concept of intrinsically disordered proteins robustly entered protein science in the mid-1990s. Nowadays, after almost two decades of fine experimental work (based on NMR and many other techniques: circular dichroism, spectroscopy, SAXS, to name just a few [39, 40]) and an exponential growth of computational papers, the field of intrinsically disordered proteins is a mature field in pure and applied protein science. Recently, in 2016, a Gordon Conference took place covering both fundamental themes and clinical applications of intrinsically disordered proteins [41].

The focus of this paper is on the distinction, perhaps overlooked in the past literature, between intrinsically disordered proteins and proteins with intrinsically disordered regions. We used a straightforward classification, based on the presence of long disordered domains (as usually done) in conjunction with the evaluation of percentages of disordered residues. We have then re-considered the theme of how intrinsically disordered proteins are related to various diseases of complex origin.

In this paper, we simply defined IDPs as proteins with $L_d \geq 30$ and $dr \geq 30\%$. On the other hand, if a protein has a long disordered segment but less than 30% of disordered amino acids, we considered it as IDRP. It has been recently observed that proteins with one long disordered segment have a higher turn-over with respect to other proteins at least in two cases: i) the segment is at least 30 amino acids long and it is located in the N or C terminus of the protein sequence or ii) the segment is at least 40 amino acids long [42]. However, if a protein has long disordered segments but, nevertheless, is poorly disordered in the rest of the sequence it is quite hard to consider it a fully intrinsically disordered protein, since it has in many cases a well-defined three-dimensional structure.

It has been reported that about 33% of proteins from Eukarya have a long disordered segment (> 30 amino acids) [23]. Xie *et al.* in a series of three papers have described the biological functions of proteins with long disordered segments [43-45]. By using the predictor PONDR VL-XT [38], Iakoucheva *et al.* shown that 66% of cancer-related proteins and 47% of signaling human proteins have a long disordered segment [24]. These estimates have been reviewed by Cheng *et al.* that reported 79% of cancer-related proteins and 66% of signaling proteins as containing long disordered segments. Moreover, they reported that 57% of the human proteins associated to cardio-vascular diseases have at least one long disordered segment [25]. Based on this type of estimates, it has been concluded that protein disorder is involved in the emergence of many diseases [16, 17].

It is important to point out that IDRPs and IDPs, more or less explicitly, have both a long disordered segment. Therefore, in the papers cited above, IDRPs and IDPs have been considered as a unique variant of proteins. However, we argued in this paper that IDRPs and IDPs must be considered as two variants, with both different physical-chemical properties and a different functional spectrum. We have shown also that IDRPs are more similar to folded proteins and quite different from IDPs. Moreover, NDPs and IDRPs are rather uniformly spread over different functional protein-classes, whereas IDPs are particularly relevant for some specific function, such as *nucleic acid binding* and *transcription factor*.

In the paper by Cheng *et al.* [25] it has been reported that the frequency of IDPs in disease-related human proteins is higher with respect to ordered proteins from the Protein Data Bank (PDB) and with respect to proteins in the Swissprot database. In our opinion, the prevalence of disorder in disease-related human proteins with respect to proteins from PDB or the entire Swissprot database is insufficient to conclude that disorder plays a role in the development of human diseases, since we cannot exclude that the frequency of IDPs in disease-related human proteins is comparable with the frequency of IDPs in the proteome of Homo Sapiens. It is well-known that the Homo Sapiens proteome contains more IDPs that other organism. Moreover, the Swissprot and the PDB database contain proteins from many organisms, including Bacteria and Archaea, that notably contain much less disorder than Eukarya [23]. In our opinion, it is more correct to compare the frequency of IDPs in human disease-related proteins with the frequency of IDPs in the Homo Sapiens proteome. This approach has been followed previously by Pajkos *et al* [46]. They observed that the percentage of disordered residues in cancer-related datasets is not different with respect to the percentage of disordered residues in the entire Homo Sapiens proteome, suggesting that cancer-related proteins do not contain more disorder than the other human proteins [46].

**Conclusions**

IDRPs are more similar to folded non-disordered proteins than to IDPs. IDRPs and NDPs have a similar functional repertoire and probably share a lock-and-key mechanism of interaction with substrates.

IDRPs and IDPs are differently present among human disease-related proteins. IDRPs probably do not play a specific role in the development of complex diseases, since their frequency is similar in disease-related proteins and in the entire human proteome. The specific relevance of IDRPs in diseases related to calcium homeostasis is an exception that deserves further investigation. As further investigation deserve the few human proteins with extended disorder, that we have called here FRAG_IDPs. It is tempting to think that these latter proteins should display an interesting repertoire of dynamical regimes, possibly due to their connection with low-complexity motifs.

IDPs are involved in the development of several diseases, but it is important to avoid statements suggesting a generic causative role of disorder in the development of diseases. It has been reported by Paikos et al. [46] that mutations related to the onset of cancer are mainly located in regions of the proteins that are not disordered. This last observation particularly suggests that the causative role of protein disorder in complex diseases should be taken with caution. Moreover, proteins related to several diseases, such as cardio-vascular diseases and diabetes, are enriched in ordered proteins and depleted in IDPs.

In a nutshell, IDPs play a specific role in the development of several diseases, but not in all human diseases, and further studies on the specific the role of disorder in the development of diseases are required.